\documentclass[letter]{aa} 
\usepackage{graphicx}
\usepackage{txfonts}
\usepackage{natbib}
\bibpunct{(}{)}{;}{a}{}{,} 
\begin{document}
   \title{First detection of a minor merger at z$\sim$0.6}

   \author{M. Puech\inst{1,2}
          \and
          F. Hammer\inst{2}
	  \and
	  H. Flores\inst{2}
	  \and
	  B. Neichel\inst{2}
	  \and
	  Y. Yang\inst{2}
	  \and
	  M. Rodrigues\inst{2}
          }

   \offprints{mpuech@eso.org}

   \institute{ESO, Karl-Schwarzschild-Strasse 2, D-85748 Garching bei
M\"unchen, Germany
\and 
GEPI, Observatoire de Paris, CNRS, University Paris Diderot; 5 Place Jules Janssen, 92190 Meudon, France
}

   \date{Received...accepted...}

\abstract{Numerical simulations predict that minor mergers are an
  important channel for the mass assembly of galaxies. However, minor
  mergers are relatively difficult to detect using imaging, especially
  at high redshift. While such events are much less violent than major
  mergers, they can nevertheless leave several features on the
  kinematical structures of remnant galaxies which could be detected
  using 3D spectroscopy. }{We present the first direct detection of a
  minor merger in a z$\sim$0.6 galaxy. Such events could indeed be
  good candidates to explain the kinematics of perturbed rotating
  disks observed with GIRAFFE at z$\sim$0.6.}{We present photometric
  and kinematical evidence of such an event in a combined analysis of
  three-band HST/ACS imaging and VLT/GIRAFFE 2D-kinematics.}{Using
  these data, we are able to demonstrate that a minor merger of a
  relatively small satellite (mass ratio $\sim$1:18) is occurring in
  this galaxy. We also derive a total SFR of
  $\sim$21$M_\odot/yr$.}{Minor mergers could be one of the physical
  processes explaining the kinematics of perturbed rotating disks,
  which represent $\sim$25\% of emission line intermediate mass
  galaxies at z$\sim$0.6. 3D spectroscopy appears to be a very good
  tool to identify minor mergers in distant (and local) galaxies.}
 

   \maketitle
%

\section{Introduction}
Minor mergers are usually defined as the merging between two galaxies
with a mass ratio smaller than 1:3 to 1:5. Hence, during a single
minor merger event, the amount of stellar mass added to the main
progenitor remains negligible. However, cosmological simulations
predict that minor mergers are one order of magnitude more frequent
than major mergers (e.g., \citealt{Khochfar06}); Moreover, they can be
responsible for \emph{in situ} star formation episodes
\citep{Woods06,Woods07} during the interaction. Therefore, minor
mergers are thought to be an important channel for the galaxy mass
assembly, even possibly the dominant contributor for intermediate mass
galaxies \citep{Guo07}.

Contrary to major mergers, minor mergers are much less violent
dynamical processes since they do not destroy the disk of the main
progenitor. Nevertheless, minor mergers can result in several
important imprints on the morphological and kinematical structure of
the remnant: they can thicken the disk of the main progenitor by
increasing its vertical scale-length and velocity dispersion, as found
both in numerical simulations
\citep{Toth92,Walker96,Velazquez99,Bournaud05,Robertson06} and
observations \citep{Schwarzkopf00}. They are believed to play an
important role during the formation of thick disks (e.g.,
\citealt{Yoachim06}), and they could also explain several other
structural properties of spiral galaxies, such as the formation of
anti-truncated surface brightness profiles \citep{Younger07}, or
activation of bars \citep{Laine99}.

In spite of their possible cosmological importance and influence on
galaxy structure, relatively few studies have focused on minor
mergers. Statistical studies aiming at identifying an important number
of minor mergers in large photometric surveys rely on counts of
galaxies in pairs which have a large difference in a red band
magnitude, used as a proxy for the stellar mass. One of the main
difficulty encountered is then the high frequency of false pairs due
to the large number of background galaxies, which makes it necessary
to have spectroscopic confirmations for both progenitors
\citep{Woods06}. Moreover, once the minor progenitor gets too close to
the main progenitor, it becomes difficult to separate both objects
using imaging alone, especially at high redshift. Therefore, detailed
studies of minor mergers are quite rare \citep{Laine99}.

Another route for identifying on-going minor mergers may be to use 3D
spectroscopy. Using GIRAFFE at the VLT, \cite{Yang07} studied the
2D-kinematical properties of a complete sample of 63 z$\sim$0.6
galaxies, which are representative of $M_{stellar}\geq$1.5
10$^{10}M_{\odot}$ emission line ($EW_0([OII])\geq$ 15\AA) galaxies.
Interestingly, they have classified 25\% of these objects as
``perturbed rotating'' galaxies (PR): their kinematics shows all the
features of a rotating disk, but with a peak in the velocity
dispersion map shifted away from the dynamical center. These local
perturbations cannot be reproduced from the rotation patterns, and
thus cannot be attributed to rotation \citep{Yang07}. They might
illustrate a possible minor merger event, which does not affect the
disk stability, but has an expected signature in the $\sigma$ map.

Other studies have also revealed the existence of morphologically
and/or kinematically disturbed galaxies at higher redshifts.
\cite{Forster06} have obtained 3D spectroscopy of several z$\sim$2
galaxies among which some also show off-centered $\sigma$ peaks in
relatively well-ordered VFs. However, all these studies (including
GIRAFFE) are seeing limited, and the physical process responsible for
such off-centered $\sigma$ peaks remains unclear, even when using
adaptive optics \citep{Law07}. In this \emph{letter}, we present what
is, to our knowledge, the first direct evidence for the detection of a
minor merger at high redshift. This object (J033226.23-274222.8) has
been classified as a PR by \cite{Yang07} and is thus an ideal case to
test the possibility that minor mergers could be a viable process
explaining PR kinematics. This \emph{letter} is organized as follows:
Sect. 1 presents the general properties of J033226.23-274222.8; Sect.
2 details the kinematical evidences for a minor merger event in this
galaxy; Sect. 3 gives some estimates for the mass and the star
formation rate; A conclusion is given in Sect. 4. We adopt $H_0$=70
km/s/Mpc, $\Omega _M$=0.3, and $\Omega _\Lambda$=0.7, and the $AB$
magnitude system.

\section{General properties}
\begin{figure}
\centering
\includegraphics[width=7cm]{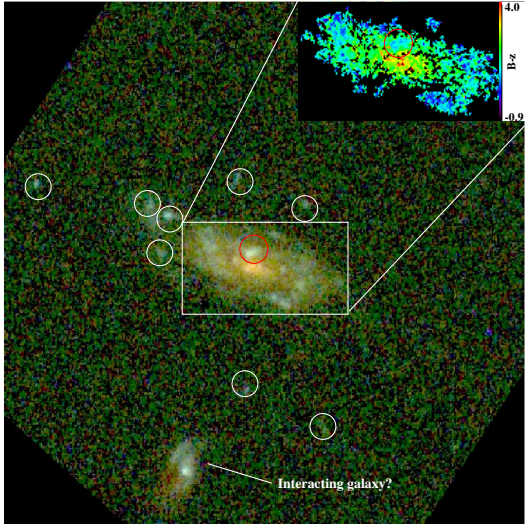}
\caption{Three bands HST/ACS imaging of J033226.23-274222.8
  ($\sim$0.03arcsec/pix). The B band is coded in blue, the V band in
  green, and the z band in red. The FoV is 16x16 arcsec$^2$; North is
  up, East is left. The contrast has been optimized by hand. White
  circles indicate the position of low surface brightness debris. The
  position of the infalling satellite is indicated by a red circle.
  \emph{Upper-right Inset:} B-z color map of J033226.23-274222.8 from
  HST/ACS imaging. The FoV is a $\sim$4.3x2.5arcsec$^2$ zoom on
  J033226.23-274222.8.}
\label{fig1}
\end{figure}

In Fig. \ref{fig1}, we show a B-V-z composite HST/ACS image, which
presents an extended view of J033226.23-274222.8. J033226.23-274222.8
is a late-type spiral galaxy with a small red central bulge (B/T=0.05)
and an extended disk (half light radius $R_{half}$=9.47kpc, see
\citealt{Neichel07}). One can note another galaxy located $\sim$5
arcsec SSE: unfortunately, we were not able to retrieve any valuable
redshift information for this galaxy, but the presence of a lot of
debris surrounding J033226.23-274222.8 suggests that these two
galaxies might be in the first stage of an interaction.

The B-z color map (see inset in Fig. \ref{fig1}) reveals a central
small bulge with a color similar to that of a passively evolved E/S0
galaxy, as well as several small ($\sim$1 kpc) blue regions located in
the disk outskirts; they have colors consistent with star-bursting
regions (see \citealt{Neichel07} for details), which suggests a
progressive inside-out build-up of the disk. At the North side of the
bulge, one can note a more extended ($\sim$2kpc) blue region (see the
red circle). This region shows a tail oriented close to West, and is
significantly redder and larger than other HII regions, which favors
the idea that this region is a small infalling satellite, rather than
another HII region. We suspect that this satellite is part of the
system of blue debris surrounding J033226.23-274222.8, which would be
infalling onto it with a trajectory as suggested by the orientation of
the tail.

\section{Kinematical evidence for a minor merger}
\begin{figure}
\centering
\includegraphics[width=6.4cm]{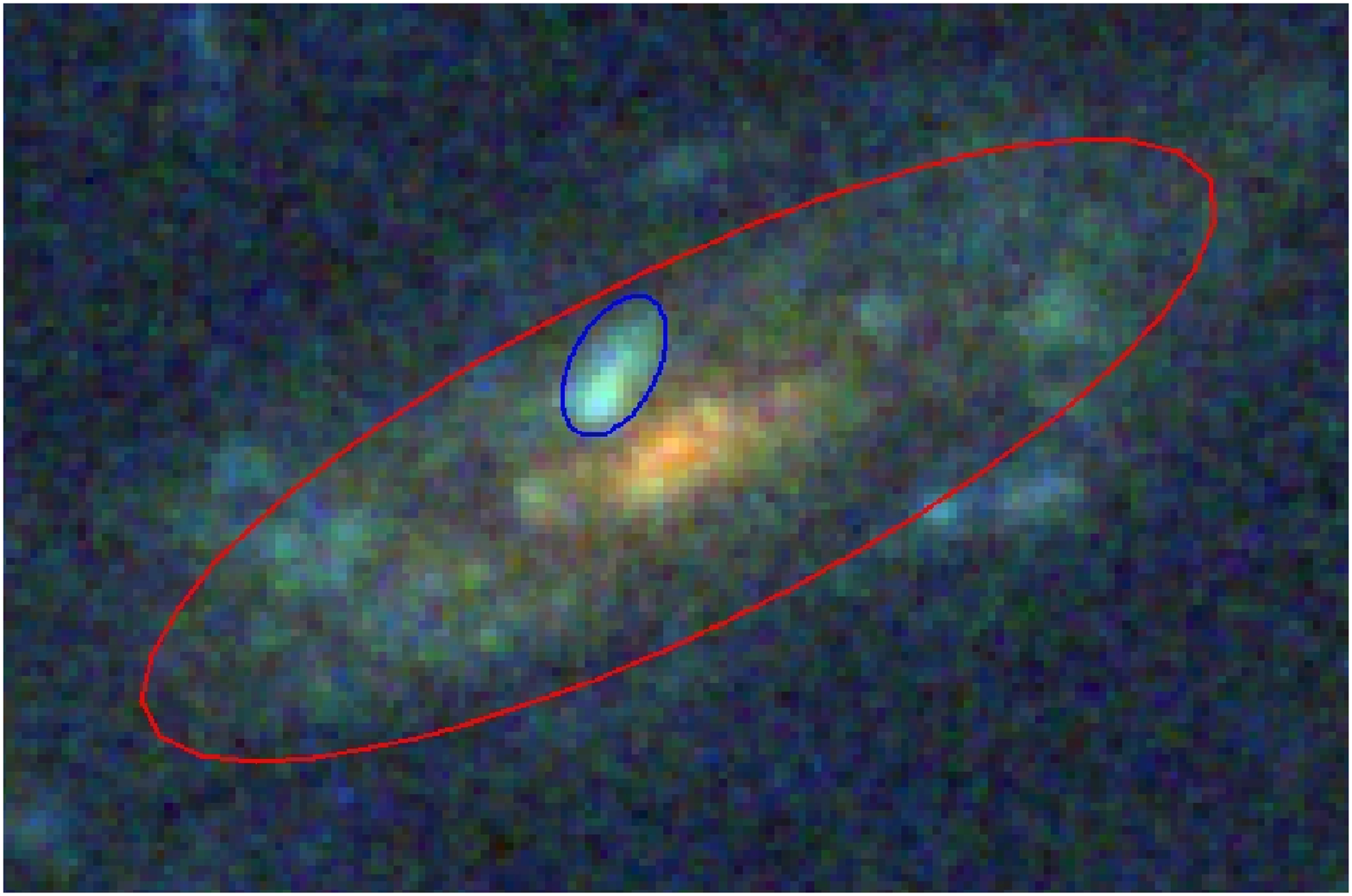}\\
\includegraphics[width=6.4cm]{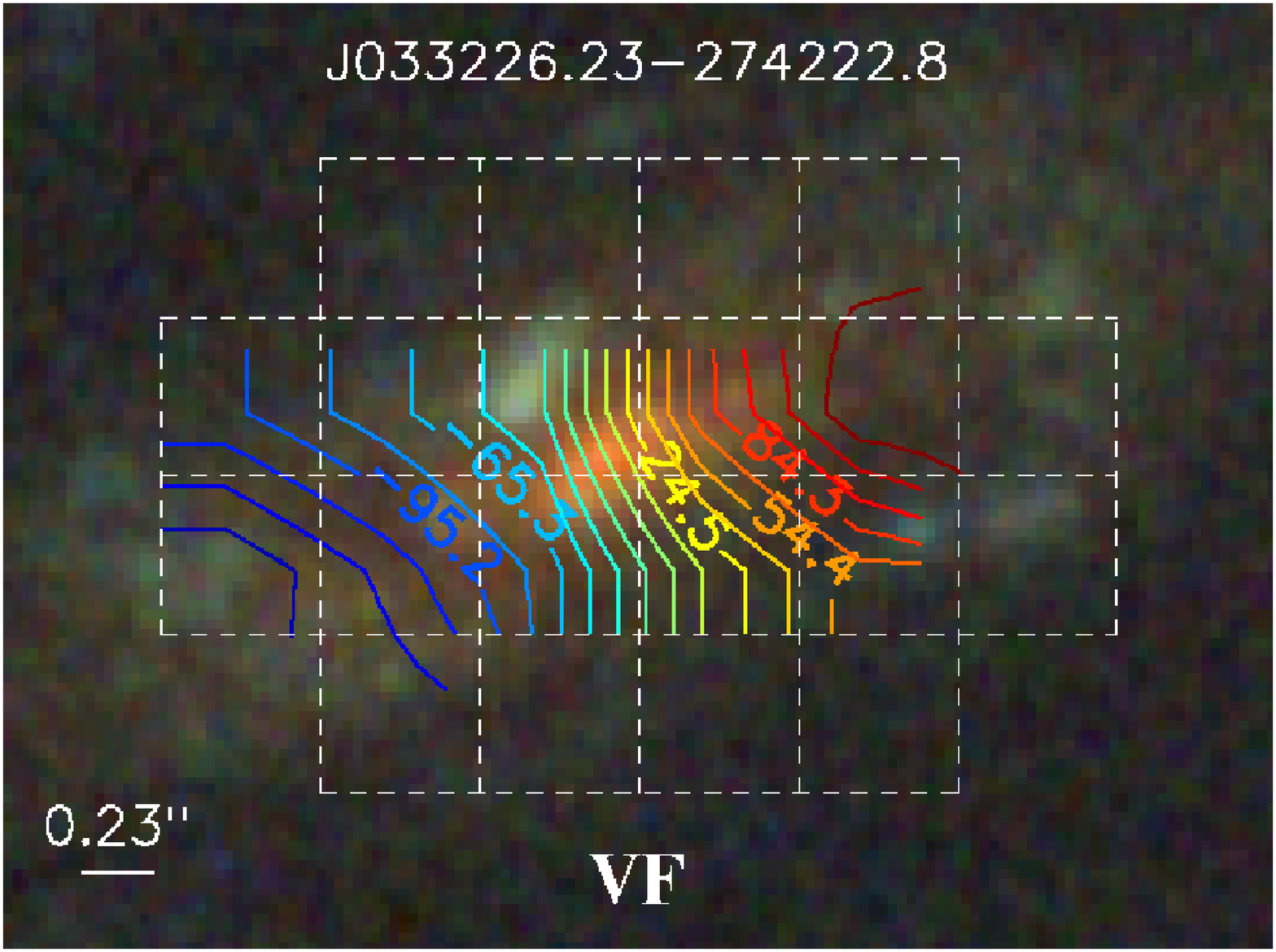}\\
\includegraphics[width=6.4cm]{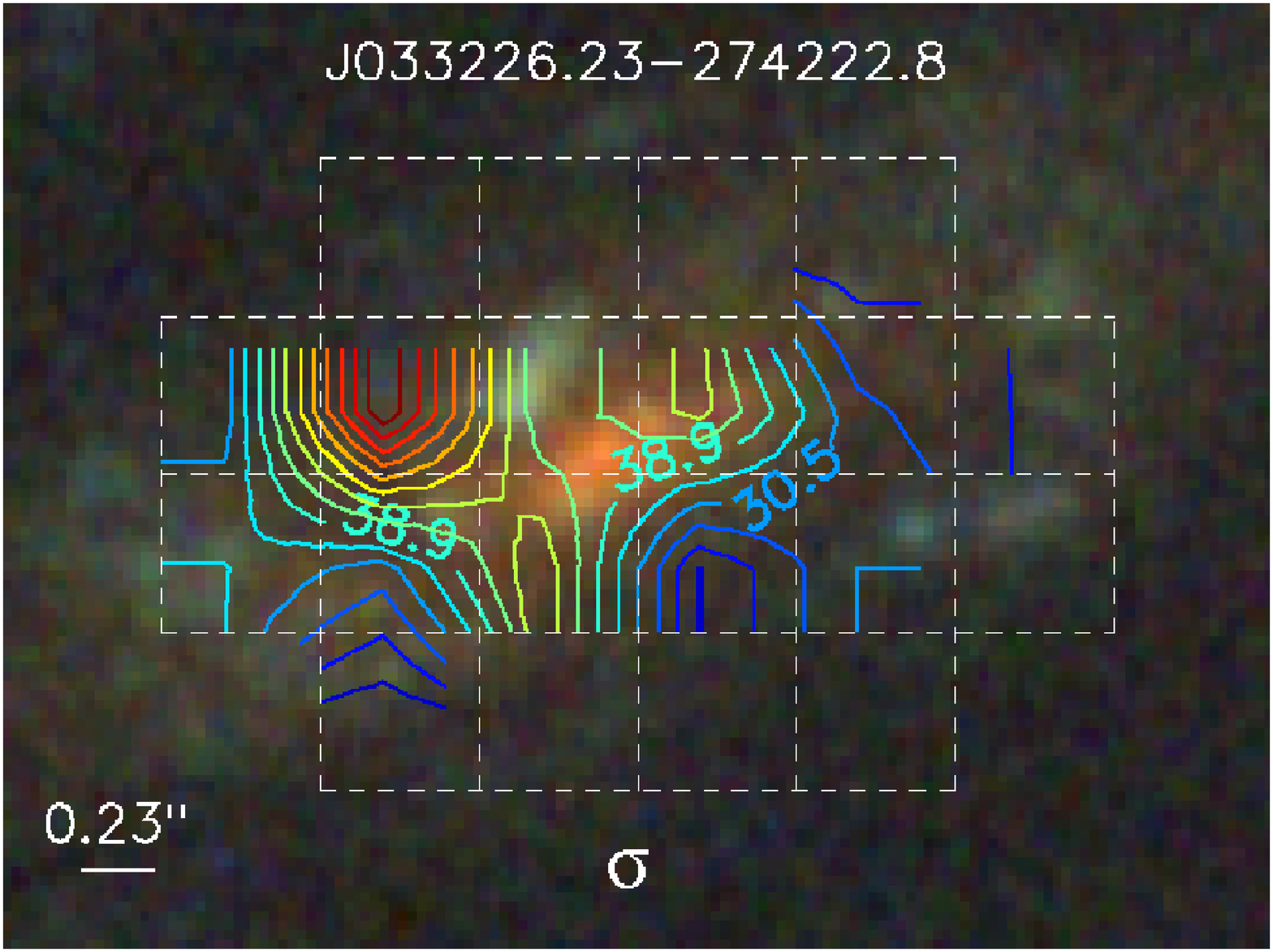}
\caption{\emph{Top panel:} Closer view of J033226.23-274222.8, which
  has been rotated to be aligned with the GIRAFFE IFU compared to Fig.
  \ref{fig1} (contrast has been optimized by hand). The two elliptical
  apertures used to estimate the stellar mass ratio between the two
  galaxies are shown in red for the main progenitor, and in blue for
  the infalling satellite (see \S~4.1). \emph{Middle and bottom
    panels:} Three bands ACS imaging of J033226.23-274222.8,
  superimposed with the GIRAFFE velocity field (\emph{middle}), or
  $\sigma$-map (\emph{bottom}). The IFU bundle is overlaid in white
  dashed lines (0.52 arcsec/pix). Note that a S/N=3 threshold is used
  to limit measurement uncertainties, which explains why kinematical
  data do not extend over the whole IFU FoV. Contours range
  approximatly from -155 to 130km/s with a 15km/s spacing for the VF,
  and from 14 to 73km/s with a 4km/s spacing for the velocity
  dispersion map.}
\label{fig3}
\end{figure}

In Fig. \ref{fig3}, we have superimposed the VF and the $\sigma$-map
derived from GIRAFFE observations \citep{Yang07} on the B-V-z
composite ACS image. Aligning ground-based IFU data with space imaging
requires accurate astrometries. In ACS images, astrometry is derived
using the GSC2 catalog, with a residual rms uncertainty on both Ra and
Dec of 0.12
arcsec\footnote{http://archive.stsci.edu/pub/hlsp/goods/v1/h\_goods\_v1.0\_rdm.html\#4.0}.
On the other hand, GIRAFFE observations are prepared using the USNO
catalog (see \citealt{Flores06}). During observations, 4 FACB stars
are used to align the GIRAFFE plate (where IFUs are positioned), with
the sky coordinate system: systematical offsets between the USNO
catalog and the observational system are then minimized. The random
residual error on this correction is estimated to be less than 0.15
arcsec rms \citep{Pasquini04}. Using Skycat on the original z-band
large field image (corresponding to the GOODS section \#34), we picked
astrometry for the 11 stars common to both GSC2 and USNO catalogs. We
used them to derive the mean offset between the two catalogs, and
correct the position of the IFU relative to ACS images: this
correction is $\sim$-0.11 arcsec in RA, and $\sim$0.04 arcsec in Dec.
Combining the rms residual uncertainties from both catalogs, we derive
a conservative $\sim$0.23 arcsec residual random uncertainty on the
position of the IFU relatively to ACS images (see the scale-bars in
the bottom-left corners), i.e., close to half a GIRAFFE pixel (0.52
arcsec/pix).

Figure \ref{fig3} shows a regular VF with no apparent perturbations,
especially close to the infalling satellite. However, given the
limited spatial resolution of GIRAFFE, only very large-scale
deviations are detectable and we cannot exclude the presence of
non-detected small scale perturbations in the VF. We have measured the
photometric $PA$ using Sextractor \citep{Bertin96} on the ACS z-band
image, and have found $PA_{phot}\sim$22{\bf $\pm2$}deg (relative to
the right side of Fig. \ref{fig3}, counterclockwise). This is very
close to the outer dynamical axis (simply defined as the line
connecting $V_{max}$ and $V_{min}$), with $PA_{kin}\sim$18{\bf
  $\pm9$}deg. We also note a progressive twist in the isovelocity
contours, which might indicate a possible warp in the gas disk. Warps
are a common feature of HI disks, although at much larger radii
\citep{Binney92}, but are also observed in a very large fraction of
stellar disks \citep{Reshetnikov98}. Interestingly,
\cite{Schwarzkopf01} found that tidal interactions and minor mergers
contribute considerably to the formation and size of warps. However,
we find that spatial resolution issues could account for at least part
of this twist. Of note, using the IRAF ellipse task, we find a
progressive twist of the (z-band) isophotal ellipses between
$R_{half}$ and 2$R_{half}$, from $\sim$30deg to $\sim$20deg,
respectively. Thus, it remains unclear whether or not this twist truly
reflects a warp.

At the GIRAFFE spatial resolution (seeing $\sim$0.8 arcsec),
undisturbed rotating disks should present a peak in their
$\sigma$-maps, close to the dynamical center. Indeed, the velocity
dispersion measured within one GIRAFFE pixel is the convolution
between random motions occurring at small spatial scales, and
integrated larger scale motions (rotation) which enlarge the emission
line profile. Thus, because of the rising part of the rotation curve
around the dynamical center, one expects this region to be dominated
by the latter. This translates into a peak in the 2D $\sigma$-maps of
regular rotating disks, located at the center of rotation
\citep{Yang07}. In the case of J033226.23-274222.8, because of the
relatively coarse spatial resolution of GIRAFFE, this central peak is
spread over two pixels located around the red bulge (see Fig.
\ref{fig3}).

\begin{figure}
\centering
\includegraphics[width=9cm]{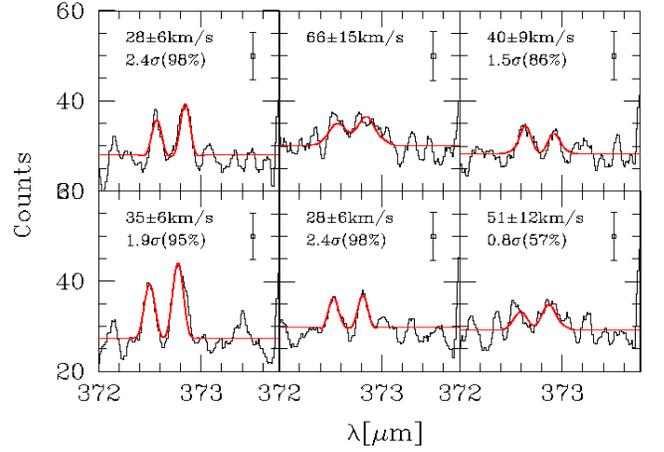}
\caption{GIRAFFE spectra in the IFU pixels surrounding the peak of the
  $\sigma$ map (here in the upper-middle panel). Observed
  sky-subtracted spectra are plotted in black, while the [OII]
  emission line fits are in red. The corresponding $\sigma$ and its
  associated uncertainty are indicated in each pixel, together with
  the level of significance of the difference with the $\sigma$ peak.
  The probability that the difference is significant is indicated into
  parenthesis (see text). Spectra have been smoothed for ease of
  visualization. A typical 1-$\sigma$ error bar due to photon noise is
  represented in each upper-right corner.}
\label{fig4}
\end{figure}

As noted in the introduction, 25\% of the GIRAFFE galaxies show
regular rotating VFs but do not show such a $\sigma$ peak at their
dynamical center. Instead, the $\sigma$ peak is found in the disk
region, as observed in the case of J033226.23-274222.8, with a peak of
66km/s located one GIRAFFE pixel left of the infalling satellite. In
Fig. \ref{fig4}, we plot the corresponding IFU spectrum and those in
adjacent pixels. One can derive the level of significance of the
difference between the peak and the dispersion measured in the
neighbor pixels using $(\sigma_{peak}-\sigma_{neighboor})/\sqrt(\delta
\sigma_{peak}^2 + \delta \sigma_{neighboor}^2)$, where $\delta \sigma$
is the uncertainty associated with $\sigma$. In doing so, we assume a
one-sided Gaussian distribution for the relative uncertainty on the
$\sigma$ measurement (see \citealt{Yang07}). A high level of
significance is found for the difference between the peak and all the
surrounding pixels. We also do not see any special variation of the
standard deviation of the fit in this pixel compared with neighboring
ones. Thus, it is unlikely that the $\sigma$ peak results from
artefacts due to S/N effects or to sky residuals.

Such an offset between the \emph{gas} $\sigma$ peak and the
\emph{stellar} mass distribution of the infalling satellite could
simply result from a decoupling between stars and gas, as observed in
numerical simulations of such interactions (e.g.,
\citealt{Hernquist95}). In this context, the $\sigma$ peak would
correspond to the location of the shock between the gas stripped away
from the infalling satellite and the gas of the main progenitor, which
would lead to an increase of $\sigma$, as observed numerically (e.g.,
\citealt{Robertson06}). Noteworthy, simulations of giant molecular
clouds show how shocks can generate a velocity dispersion
$\sigma[km/s] \sim L[pc]^{1/2}$ \citep{Bonnell06}, where $L$ is the
spatial scale of interest (i.e., at which $\sigma$ is measured).
Assuming that this relation can be extrapolated to galactic scales,
one finds that at the scale of a GIRAFFE pixel ($\sim$3.6kpc), the
shock-generated $\sigma$ would be $\sim$60km/s, which is consistent
with the peak value.

\section{SFR and mass estimates}

\subsection{Stellar and dynamical mass}
In \cite{Puech07b}, we derive a rotational velocity $V_{rot}=200$km/s,
corrected for inclination and spatial resolution effects. Following
\cite{Puech07a}, we estimate the V/$\sigma$ ratio in the disk to be
$\sim$8.8$\pm$1.8, compatible with a thin late-type rotating disk (see
also \citealt{Fathi07}). We can thus assume full rotational support,
which leads to a total dynamical mass within the optical radius
($\sim2R_{half}$), $Log(M_{dyn}/M_\odot)\sim11$. It is however not
possible to derive a direct estimate of $M_{dyn}$ for the accreted
object because the $\sigma$ peak probably traces shock heating rather
than gravitational motions, since it does not correspond to the
underlying mass distribution (see discussion in \citealt{Colina05}).

On the other hand, it is possible to derive a rough estimate of the
\emph{stellar} mass ratio between the two progenitors, by defining two
elliptical apertures on the z-band image, the reddest image at our
disposal. For the main progenitor, we used directly the elliptical
aperture returned by Sextractor. For the infalling galaxy, we defined
an elliptical aperture by hand, guided by the B-V-z image (see Fig.
\ref{fig3}). We then estimated the stellar mass ratio between the two
progenitors by simply deriving the ratio between the counts obtained
within the respective apertures. Finally, we corrected this by the V-z
color difference between the two progenitors ($\sim$0.24), which we
approximate to the u-g color at rest. Using the stellar mass
calibration given by \cite{Bell03}, we find a final mass ratio of
$\sim$1:18.

Following \cite{Ravikumar07}, the stellar mass for the two progenitors
is $Log(M_{stellar}/M_\odot)=10.72$, which leads to
$Log(M_{stellar}/M_\odot)\sim10.7$ and $Log(M_{stellar}/M_\odot)=9.5$
for the main and minor progenitors respectively. Of note, this
correspond for the main progenitor to a $M_{stellar}/M_{dyn}\sim$0.5,
which is in the range of values measured by \cite{Conselice05}.
Moreover, J033226.23-274222.8 follows the relation
$M_{dyn}=10^{-2.4}M_{stellar}^{1.25}$ found by \cite{Rettura06}:
assuming that this relation can be extended to galaxies having
$Log(M{stellar}/M_\odot)\leq$10, a stellar mass ratio of 1:18 leads to
a dynamical mass ratio of $\sim$1:37. We are thus quite confident that
we are in the presence of the accretion of a relatively small
satellite.

\subsection{Star formation rate}
We first use the 24$\mu$m MIPS DR-3 data to derive the dust-obscured
SFR$_{IR}$. At z=0.6679, this flux roughly corresponds to a rest-frame
15$\mu m$ flux: we can thus use the \cite{Chary01} calibration to
derive the total IR luminosity $L_{IR}$, and then use the classical
\cite{Kennicutt98} calibration between $L_{IR}$ and SFR$_{IR}$. We
find SFR$_{IR}=17.5M_\odot/yr$, corresponding to $f_{24\mu m}=91.7\mu
Jy$. To determine the unobscured SFR$_{UV}$, we used the UV
calibration of \cite{Kennicutt98}, using the rest-frame 2800\AA~
luminosity. We find $M_{2800}=-20.12$, leading to SFR$_{UV}=3.2M_\odot
/ yr$. We thus find a total SFR$_{tot}$=SFR$_{UV}$+SFR$_{IR}\sim20.7
M_\odot / yr$. Assuming a constant SFR during the disk build-up, and
neglecting the contribution of the bulge to the total stellar mass
(B/T=0.05), we can roughly estimate a timescale for the formation of
the disk using $M_{stellar}$/SFR$_{tot}\sim$2.5Gyr.

We used the B-z colormap to define star-bursting regions in the disk
and the satellite as those having B-z$<$1 \citep{Neichel07}. We find
that the satellite star-bursting regions represent $\sim$15\% of the
B-band flux located in all star-bursting regions of the disk
(satellite excluded). Moreover, the ratio between the obscured and
unobscured SFR is $\sim$5.5, which is lower than the median ratio
found in 0.4$\leq$z$\leq$1 star-burst galaxies ($\sim$13, see, e.g.,
\citealt{Hammer05}): assuming there are no completely dust-obscured
star forming regions, this leads us to a contribution of HII
star-bursting regions to SFR$_{tot}$ of $\sim$85\%, corresponding to
$\sim$17.6$M_\odot/yr$. Therefore, the SFR enhancement due to the
satellite would be $\sim$3.1$M_\odot/yr$, which is comparable to what
is found in observations \citep{Woods07}.

\section{Conclusion}
In this \emph{letter}, we have presented the first detection of a
minor merger at high redshift (z=0.6679). Using ACS imaging, we
identified an infalling satellite onto a late-type spiral. Gas
kinematics reveals a velocity field with no large-scale perturbations,
except a twist in the isovelocities which might indicate a possible
warp of the gaseous disk. Such a warp is also seen in the stellar
disk, using photometric isophotes. The gas $\sigma$-map shows a peak
off-centered from the dynamical center, and close to the infalling
satellite. Such a peak cannot be attributed to rotation and is
interpreted as the signature of shock-heated gas during the collision.
We derive a total SFR$\sim$21$M_\odot/yr$ and a rough stellar mass
ratio estimate $\sim$1:18, in agreement with this kind of events.
Therefore, J033226.23-274222.8 seems to be the first detection of a
minor merger at z$\sim0.6$, and more generally at high redshift. This
study demonstrates that using 3D spectroscopy appears to be a
promising tool to identify minor mergers, both in nearby and distant
galaxies. Finally, minor mergers are a good candidate to explain the
kinematics observed in perturbed rotating galaxies, which account for
25\% of emission line intermediate mass galaxies at z$\sim$0.6.

\begin{acknowledgements}
The authors wish to thank an anonymous referee for useful comments
and suggestions.
\end{acknowledgements}

\end{document}